# Transverse hypercrystals formed by periodically modulated phonon-polaritons

**Authors:** Hanan Herzig Sheinfux[1], Minwoo Jung[2], Lorenzo Orsini[1], Matteo Ceccanti[1], Aditya Mahalanabish[2], Daniel Martinez-Cercós [1], Iacopo Torre[1], David Barcons Ruiz[1], Eli Janzen[3], James H. Edgar[3], Valerio Pruneri[1], Gennady Shvets[2], Frank H.L. Koppens[1,4]

[1]ICFO-Institut de Ciencies Fotoniques, 08860 Castelldefels (Barcelona), Spain

[2] School of Applied and Engineering Physics, Cornell University, Ithaca, New York 14853, USA

[3]Tim Taylor Department of Chemical Engineering, Kansas State University, Durland Hall, Manhattan, KS 66506-5102, USA

[4]ICREA-Institució Catalana de Recerca i Estudis Avançats, 08010 Barcelona, Spain

**Photonic crystals and metamaterials are two overarching paradigms for manipulating light. Combining the two approaches leads to hypercrystals: hyperbolic dispersion metamaterials that undergo periodic modulation and mix photonic-crystal-like aspects with hyperbolic dispersion physics. So far, there has been limited experimental realization of hypercrystals due to various technical and design constraints. Here, we create nanoscale hypercrystals with lattice constants ranging from 25 nm to 160 nm and measure their collective Bloch modes and dispersion with scattering nearfield microscopy. We demonstrate for the first time dispersion features such as negative group velocity, indicative of bandfolding, and signatures of sharp density of states peaks, expected for hypercrystals (and not for ordinary polaritonic crystals). These density peaks connect our findings to the theoretical prediction of an extremely rich hypercrystal bandstructure emerging even in geometrically simple lattices. These features make hypercrystals both fundamentally interesting, as well as of potential use to engineering nanoscale light-matter interactions.**

Drawing inspiration from natural crystalline systems, photonic crystals are dielectric structures which change periodically on a scale typically comparable to the wavelength of light[1–3]. Light propagating in a photonic crystal undergoes multiple scattering events and forms collective Bloch modes, analogously to electrons in an atomic crystal. By engineering the periodic modulation, the photonic bandstructure can be extensively tailored and features such as bandgaps and density of states (DOS) peaks can be induced or manipulated, enabling a large range of applications, such as nanophotonic confinement, guiding or lasing[4,5]. In a further analogue to crystalline matter, photonic crystals can also be made topological and can exhibit properties such as protected transport[6].

The same paradigm has been applied at the nanoscale using polaritonic excitations[7–9] with a wavelength that is orders of magnitude shorter than the vacuum wavelength[10]. Specifically, two material platforms have been used to realize polaritonic crystals. Phonon polaritons (PhPs) in a hexagonal boron nitride (hBN) flake exhibited a crystalline bandstructure when the flake was periodically perforated[11,12]. However, the damage incurred in patterning the hBN adversely affects the PhP lifetime, damping the collective modes. This restricts the possibility to directly observe Bloch modes and characterize their momentum, and poses a practical limit on reduction in the lattice period from ~600 nm (as in [12]) to the nanoscale (i.e. to below 100 nm). Another polaritonic crystal system was realized with graphene plasmon polaritons, using periodic electrostatically-induced doping with an 85 nm period. At cryogenic temperatures, clear collective graphene plasmon polariton modes were produced[13], with a striking dependence of the amplitude and wavelength on the excitation direction[14]. However, extending these ideas to generate richer and more complicated bandstructures, e.g. topological graphene plasmon crystals[15], is technologically challenging.

Beyond the miniaturization of photonic crystals, is there any new physics unique to periodically modulated polaritons? Periodically modulated hyperbolic media were first proposed[16] and realized[17–19] in the context of metamaterials[20,21], where they are known as 'hypercrystals'. In ordinary photonic crystals light has a natural length scale, the wavelength, but in a hypercrystal, no such length scale exists. At the same frequency many PhP modes exist, each mode with its own wavelength[20,22–26].

For example, consider a slab of material with an anisotropic permittivity $\bar{\bar{\epsilon}} = \text{diag}(\epsilon_x, \epsilon_y, \epsilon_z)$, extended into x-y and with a finite thickness $t$. Modes propagating in the x-direction are governed by the anisotropic Helmholtz equation: $\partial_z^2 E_x(x,z) = \epsilon_z k_0^2 E_x(x,z) - \frac{\epsilon_z(\omega;z)}{\epsilon_x(\omega;z)} \partial_x^2 E_x(x,z)$, with $\omega$ the angular frequency and $E_x$ being the x component of the electric field. Ordinarily, $k_0 \equiv \frac{\omega}{c}$ sets a cut-off on the maximum wavevector of any propagating plane wave excitation. If the x-component of the wavevector exceeds this cut-off, the right side of this equation becomes negative and the excitation must decay exponentially. But if the permittivity components have opposite signs, such that $-\frac{\epsilon_z(\omega;z)}{\epsilon_x(\omega;z)}$ is a (mostly) real positive number, there is no such limitation. The resulting modes, at a given $\omega$, are similar to those of a dielectric waveguide slab (see detail in SI section 1), except that the number of modes is independent of the thickness of the slab and can be very large (in principle, infinite).

If this hyperbolic media slab is now periodically modulated in the x-y direction, we designate it as a transverse hypercrystal (THC). Since the wavelengths of the slab modes are not simple harmonics of each other, each mode has its own length scale. Therefore, even if no coupling between various PhP modes exists, applying a periodic potential to each mode will result in a separate bandstructure for each mode. If any degree of intermodal coupling does exist, it will make the bandstructure richer and more complex. Most notably, this richness of the bandstructure can be expected to occur even if the underlying crystal structure is trivially simple. A 1D grating or a 2D square lattice should generate an extremely complex bandstructure.

The emergence of complexity from simple geometries is one of the most intriguing themes of modern physics. It famously occurs in the context of chaos theory and of Hofstadter's butterfly in condensed matter. However, to the best of our knowledge, emergent complexity has not been realized, or even theoretically predicted, in the context of nanophotonics. In the specific case of hypercrystals, this is because in all experimental realizations[11,12,18,27] technical details in the design[28] greatly diminish the experimental signature of higher order modes.

Here, we realize THCs with strong multimodal interaction and experimentally study their collective modes and dispersion. Our calculations predict, for the first time, that even geometrically simple lattices can support an incredibly rich spectrum, composed of an array of overlapping bands. Moreover, we experimentally demonstrate spectral signatures of this rich bandstructure, which defy conventional single mode analysis of polaritonic crystals. Experimentally, our THCs are made by placing hBN on a patterned metallic substrate. The typical THC lattice period is 100-160 nm, but we also observe collective Bloch modes in lattices with an unprecedentedly small period of 25 nm. We extract the Bloch wavevector of these modes to demonstrate dispersion features such as group velocity sign switching, previously unobserved at the nanoscale. In addition, we measure extremely narrow bandwidth response in the hypercrystal, associated with sharp DOS peaks. These DOS peaks are not expected to occur in a single mode analogue of the same lattice and are therefore indicative of multimodal interaction. This links to our theoretical prediction of a rich THC bandstructure, with hundreds of coupled and interacting bands.

The experimental platform is an extension of the technique developed in [29] and consists of an hBN flake on top of an array of closely packed gold nanoislands (see Fig. 1a). The nanoisland array is produced by a focused ion beam (FIB) milling of an ultraflat layer of gold[30]. After milling, an isotopically pure hBN flake is mechanically exfoliated and transferred on top of the patterned substrate using standard techniques (see SI section 3). The hBN flake remains flat, suspended between the nanoislands, and does not conform to the substrate topography. In addition, the hBN flake maintains an as-exfoliated high quality (relatively low PhP absorption). Below we shall refer to four hBN lattice devices, NSQ1, NSQ2 and HSQ1, which are composed of square nanoislands in a square lattice and NGR1 being a 1D lattice. Each device is characterized by a period, $p$, metal nanoisland size, $w$, and hBN flake thickness, $t$ (see schematic in Fig. 1a,1b and further detail on fabrication in the SI).

The PhP modes in an hBN flake are analogous to the TM polarized modes of a dielectric waveguide slab, with the exception that due to hyperbolicity, many equal-frequency modes coexist in the slab regardless of its thickness. In further analog to a dielectric slab, the PhP modes depend on the permittivity of the substrate and superstrate through the phase of the reflection coefficient from the interfaces[31,32]. Changing the substrate modifies both the momentum and the spatial profile of the PhP modes and induces intermodal coupling. To illustrate this, Fig. 1c shows a single mode launched and reflected from the interface at which the substrate changes between dielectric and metallic. The reflected and transmitted field profile shows strong ray-like components, generated by the interference of higher order modes[33]. This reflection and intermodal coupling are particularly strong when the substrate changes from metallic to dielectric (and much less so for a purely dielectric substrate[27]) and strongly depend on the sharpness of the edges, since smoother edges are expected to introduce weaker intermodal coupling.

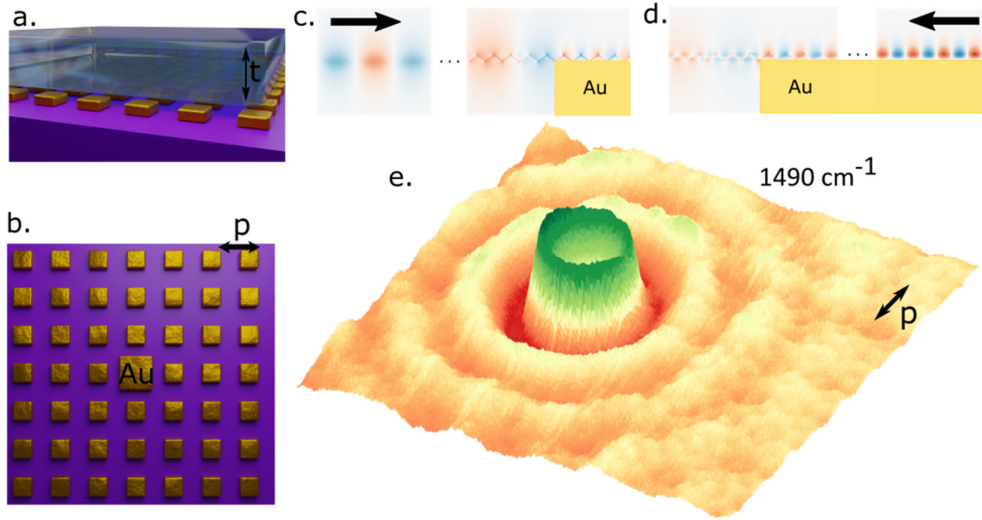

**Fig. 1, Phonon polaritonic hypercrystal composition and collective modes. a.** Schematic side and **b.** top view of a hypercrystal, including an intentional structural defect. **c.** Simulation of multimodal coupling, where the lowest order mode is launched from the dielectric substrate side and reflected from a sharp metallic corner. The mode is launched far from the metallic corner (as indicated by '…'). Multiple order modes are reflected and transmitted from the metallic corner. **d.** Same as c, but for the mode launched from the metal side. **e.** Experimental (nearfield) signal from device NSQ1, taken near the structural defect at $f = 1490 \text{cm}^{-1}$. The lattice period is $p = 100$nm.

The collective PhP modes are measured with a mid-IR scattering nearfield optical microscope (sSNOM), a scanning probe technique which can image the nearfield response with typical resolution of ~20nm. This nearfield response is related to the local photonic DOS (see SI). As the sSNOM tip is scanned, it both launches and picks up polaritons in the underlying sample. If these polaritons are reflected by structural features in the sample (i.e. from the edge of the lattice or a structural defect), the local density of states increases or decreases depending on whether the reflected polariton experiences constructive or destructive interference with the tip-launched polariton. For example, Fig. 1e shows the measured sSNOM signal of device NSQ1 ($p = 100$nm, $w = 25$nm, $t=23$nm) in the vicinity of a structural defect (four times larger than the other nanoislands). This defect reflects tip-launched polaritons[34], forming circular rings of constructive\destructive interference. At the same time, the sSNOM signal also probes the local DOS of the waves inside the unit cell, which can be thought of as the PhP Bloch function[11,27] (see also additional scans in SI section 5.1). The same ideas can also be applied to much shorter lattice periods, for example, down to 25 nm (see SI section 5.2).

In what follows, we focus on the larger period devices, such as HSQ1 ($p = 120$nm, $w = 85$nm, $t = 22$nm) where precise fabrication and characterization are possible. Fig. 2 shows a sequence of nearfield maps of HSQ1 taken at sequentially increasing frequencies, where we can identify several spectral regimes: **I.** For lower frequencies, collective modes are predominantly along the $\Gamma - M$ direction (at 45° to the lattice vectors). The wavelength of these collective modes decreases with increasing frequency, corresponding to positive group velocity. Notably, these modes are relatively long lived and the fringes extend over a length comparable to PhPs on the unpatterned metal, indicating that geometric spread of the wavefront is the dominant signal decay mechanism (see SI section 5.3). **II.** As the frequency increases, collective modes disappear (not shown in the figure). However, the absence of collective modes cannot be unambiguously attributed to a bandgap-like feature. If the Bloch wavevector increases, coupling to the sSNOM tip becomes increasingly inefficient. Likewise, if the wavevector decreases significantly, the wavelength becomes larger than the patterned region and it becomes impossible to discern the collective mode (as happens, for example, for frequencies below 1460cm$^{-1}$). **III.** Collective modes reappear, manifesting exclusively along the $\Gamma - X$ direction. Importantly, the collective modes now exhibit negative group velocity, with the wavelength increases with increasing frequency. This type of group velocity flip was not observed before in any comparable nanoscale system. **IV.** Collective modes again disappear and the entire lattice shows a strong uniform response, which could be attributed to a relatively flat band. **V.** Positive group velocity polaritons reappear, oriented along the $\Gamma - M$ direction.

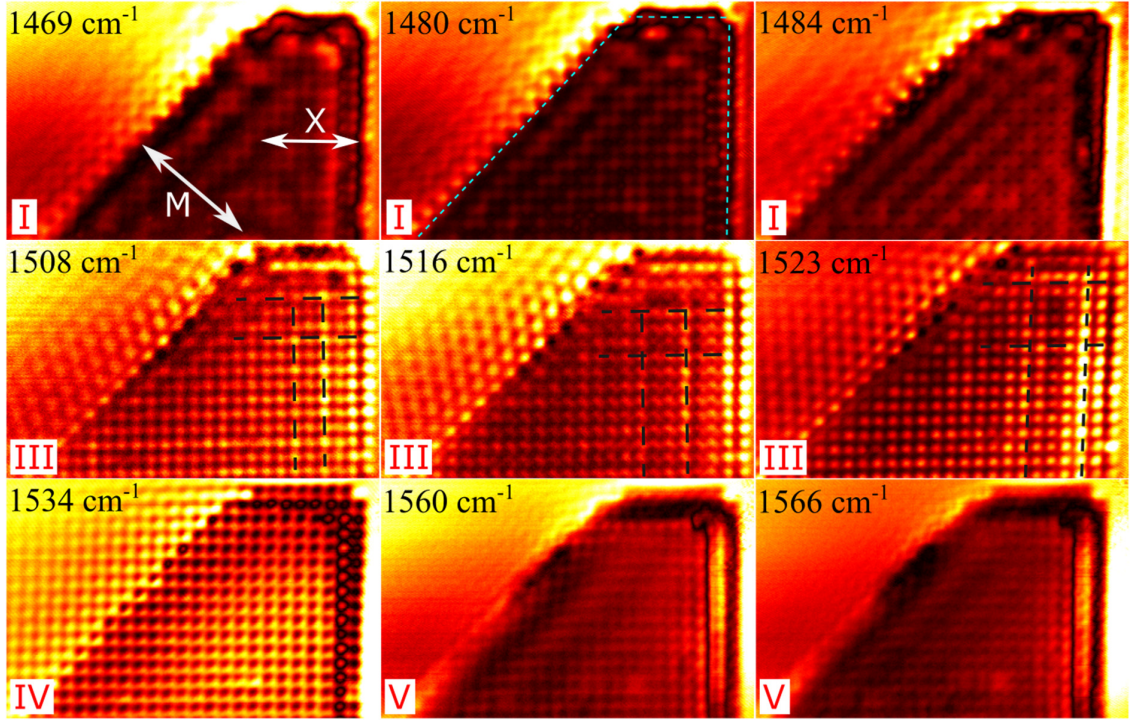

**Fig. 2, Evolution of the collective mode with frequency.** The array of homodyne sSNOM measurements (4$^{th}$ harmonic) corresponding to the frequencies indicated on each subplot. Black dashed lines are guides to the eye, showing the location of the fringe peaks. White double arrows indicate the $\Gamma - M$ and $\Gamma - X$ directions. Bright blue dashed lines indicate location of the lattice edge. Roman numerals indicate the dispersion regime corresponding to each figure: In regime I, the group velocity is positive (wavelength decreases with frequency), in regime III, it is negative and in regime V it is positive again. Scans in regimes II and IV show no collective modes (regime II is not shown in the main text). Scale bar in the top left is 500 nm and is applicable to all subfigures.

We can compare these results with the simulated dispersion of device HSQ1, for three cases: **(1)** A simplified case, where we artificially assume that only a single mode exists. That is, the bandstructure of a 2D photonic crystal slab which is similar to our device, but only has a single mode per frequency. The local refractive index of this mode equals the effective refractive index of the lowest order PhP mode at the same frequency and above the same substrate. The calculated bandstructure shown in Fig. 3a is typical for a 2D square crystal and, in the spectral region of interest, shows a few bands and a single bandgap. **(2)** The full calculation, shown in Fig. 3b, results in a THC bandstructure so dense that it almost seems continuous, consisting of hundreds of bands, an array of DOS singularities (scars) and no bandgaps. It bears only a trace of resemblance to the single mode bandstructure. The complexity of this bandstructure is remarkable and, to the best of our knowledge, has not been previously predicted.

**(3)** A single folded version of the full calculation is shown in Fig. 3c, where by single folding we mean that the PhP momentum can change by up to one lattice vector (see precise definition in SI section 2.3), in contrast with full bandstructure folding (as in Fig. 3a and Fig. 3b), where the collective mode can change any number of lattice vectors. The single folding better represents the experimental reality where the sSNOM tip couples much more efficiently to momentum components in the first Brillouin zone and where absorption limits the number of times a given momentum component can interact with the underlying lattice.

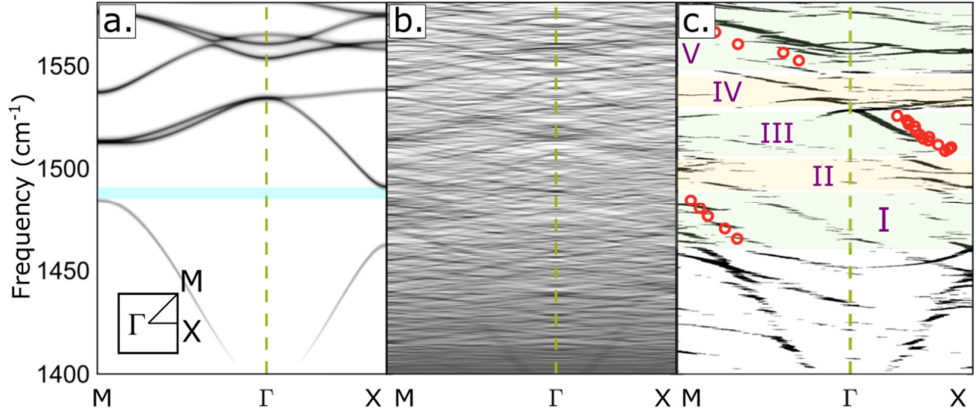

**Fig. 3, Hypercrystal bandstructure. a.** Calculated bandstructure (energy loss function) of a single mode analogue of device HSQ1. The blue highlight shows a bandgap. Inset shows the Brillouin zone structure. **b.** Calculated bandstructure of device HSQ1, including high order modes, intermodal coupling and experimental absorption. **c.** The same, but performing only a single folding of the bandstructure and including experimentally extracted momentum (red circles). Roman numerals on color-highlighted areas signify different dispersion regimes, which correspond to experimental regions.

Superimposed on Fig. 3c are the different dispersion regimes (I through V) and the PhP momenta extracted from the experiment (oriented along the $\Gamma - X$ or $\Gamma - M$). These experimental PhP momenta are in good agreement with the single folding bandstructure. They are also in good agreement with the single mode bandstructure with the notable exception that in the experiment (and in the single folded bandstructure) modes in regime III propagate predominantly along the $\Gamma - X$ direction, whereas in the single mode model, they are expected to be more pronounced in the $\Gamma - M$ direction.

Next, we study multimodal features with device NSQ2 ($p = 120$nm, $w = 25$nm), which by design exhibits a very simple single mode bandstructure: two bands separated by a single bandgap around 1520 cm$^{-1}$. The DOS that corresponds to the single mode model is therefore smooth except for the step-like drop to zero at the edge of the bandgap (the Van Hove singularities). In contrast, the bandstructure obtained from the full (multimodal) calculation is complex and segmented (see SI section 5.3), similarly to bandstructure in Fig. 3b,3c, resulting in an array of narrow-bandwidth DOS peaks. To study these experimentally, we perform a frequency sweep: repeatedly scanning along a single line, while changing the laser frequency in small (<1cm$^{-1}$) steps. The result (Fig. 4a) shows two sets of PhP fringes emanating from the defect (a single enlarged nanoisland): a low momentum branch prominent above 1420 cm$^{-1}$ and a high momentum branch which is prominent below 1420 cm$^{-1}$. These two modes originate from different order modes of the hBN slab. We analyze the dispersion of these modes in detail, similarly to what was done for device HSQ1, in the SI (section 5.3). Moreover, the frequency sweep also shows an area of relatively low signal, denoted pseudobandgap (PBG), and a few spectrally narrow peaks, such as those indicated in Fig. 4a by dashed green lines.

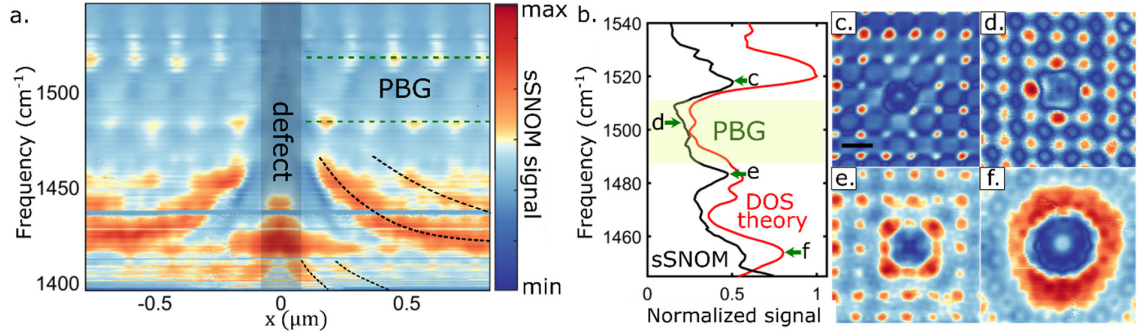

**Fig. 4, Density of states peaks. a.** Frequency sweep along a single line, crossing through an intentionally placed structural defect. The dashed black lines are guides to the eye indicating the location of polaritonic fringes emanating from the defect. The dashed green lines indicate DOS peaks which surround a pseudobandgap (PBG) spectral region of reduced DOS. Horizontal streaks, as in below 1440 cm$^{-1}$ are due to laser absorption. **b.** Integrated signal from the frequency sweep over a single unit cell, compared against the computed DOS for the experimentally accessible region of momentum space. Importantly, single mode DOS is not expected to show any peaks in this spectral region. **c.-f.** Measurements around the structural defects for frequencies near inflection points in the DOS: 1516, 1498, 1483, 1453 cm$^{-1}$, respectively. Panel f shows a Bloch mode emanating from the defect, whereas panels c-e do not.

To interpret these narrow bandwidth peaks, we integrate the signal from the frequency sweep dataset over one unit cell (far from the defect, to avoid PhP fringes). The integrated experimental signal (black curve in Fig. 4b) is in good agreement with the theoretical red curve, obtained from a weighted average of the unfolded bandstructure (weighted to account for the coupling efficiency, see detailed explanation in SI). The PBG dip around 1500 cm$^{-1}$ is not a proper bandgap (indeed, the fully folded bandstructure shows no bandgaps), it does appear to play a role similar to that of a proper bandgap and seems to originate in the bandgap of the single mode structure. In addition to the PBG, we also observe three relatively narrow signal peaks at 1453, 1483 and 1516 cm$^{-1}$. These peaks are associated with the nearfield maps in Fig. 4c-f, showing different unit cell response (Fig. 4f also shows a propagating mode). The peak at 1483cm$^{-1}$, in particular, is about 180 times narrower than the excitation frequency and another device shows even narrower peaks (see SI section 5.5). This suggests large quality factors and potential applications for hypercrystals in enhancing light-matter interactions.

These narrow bandwidth peaks (associated to sharp DOS peaks) are an important experimental signature of the multimodal nature of the bandstructure. While these narrow features appear in the experiment and in the full (multimodal) calculation, they are completely absent from the single mode picture. In fact, sharp DOS peaks in a 2D bandstructure are only expected in the presence of a saddle point in the bandstructure. In the single mode bandstructure, saddle points only appear outside the measurement range (above 1535 cm$^{-1}$). The discrepancy between single mode and multimodal is particularly evident for 1453 cm$^{-1}$. Both the experiment and the full calculation show a narrow peak around 1453 cm$^{-1}$, but this frequency corresponds to the middle of the band in the single mode calculation, so the single mode DOS should change uniform$v$7 and exhibit no peaks. These signal peaks are related to the segmented character of the bandstructure (similar to Fig. 3c, see SI section 5.4). They therefore provide a link between our measurements and the theoretical prediction of emergent complexity in a geometrically simple structure, occurring via multimodal interaction.

In conclusion, our work demonstrates a new platform for polaritonic bandstructure engineering, enabling dispersion control of nanoscale polaritons, including manipulation of group velocity and generation of DOS peaks. These DOS peaks constitute a signature of the multimodal interaction and the hyperbolic nature of the PhPs used and are suggestive of the richness of hypercrystal bandstructure, which has not been predicted in the past. The contrast between the simple geometry and the complex bandstructure which emerges in the hypercrystal naturally opens new directions for complex hypercrystals, such as the possibility of topological properties arising in hypercrystals with a non-trivial unit cell.

# Methods

Additional discussion of techniques is available in the supplementary information.

Sample fabrication

Lattices are milled by depositing an ultrathin gold layer on a dielectric substrate. In devices NSQ1, NSQ2, we used a 10nm Au with a Ti seed layer over an $SiO_2$ substrate. In devices NGR1, HSQ1, a 1nm Cu seed layer and 8nm Au were used and the substrate was either $SiO_2$ (device NGR1) or a thin SiN membrane (device HSQ1). We patterned the metallic surface with a focused ion beam microscope (Zeiss Orion) using $Ne^+$ ions (devices NSQ1, NSQ2, NGR1) or $He^+$ ions (device HSQ1), with $5\mu T$ gas pressure and a 10um aperture. Isotopic hBN flakes are then mechanically exfoliated and transferred onto the substrate with polydimethylsiloxane (PDMS) based exfoliation and transfer (X0 retention, DGL type from Gelpak) at 80C°. If deemed necessary, contact mode AFM was used to clean the surface.

Device parameters are concentrated below, for convenience.

| Device | $p$ (nm) | $t$ (nm) | $w$ (nm) | Fabrication |
|---|---|---|---|---|
| NSQ1 | 100 | 23 | 25 | Ne FIB |
| NSQ2 | 160 | 23 | 40 | Ne FIB |
| NGR1 | 25 | 3 | ~12 | Ne FIB |
| HSQ1 | 120 | 22 | 85 | He FIB |

Nearfield measurements

All measurements were performed using a commercially available scattering-type near field microscope (Neaspec), equipped with Pt coated AFM tips (Arrow NCPt from Nanoandmore, nominal diameter of 40-50nm). The laser source was a tunable quantum cascade laser (Daylight Instruments MIRcat), giving 10 to 90mW CW laser power, depending on frequency. All reported measurements were done in homodyne mode, to maximize signal to noise ratio and stability.

Theoretical analysis

All numerical calculations were made using COMSOL Multiphysics. Single mode calculations were done assuming that the local refractive index equals the effective refractive index of the lowest order PhP mode in an hBN slab at the same frequency and over an identical substrate. Multimodal calculations were done without any such assumptions. See extended detail in SI section 2.

Acknowledgements

F.H.L.K. acknowledges support by the ERC TOPONANOP under grant agreement n° 726001, the Government of Spain (FIS2016-81044; Severo Ochoa CEX2019-000910-S), Fundació Cellex, Fundació Mir-Puig, and Generalitat de Catalunya (CERCA, AGAUR, SGR 1656). Furthermore, the research leading to these results has received funding from the European Union's Horizon 2020 under grant agreement no. 881603 (Graphene flagship Core3). H.H.S. acknowledges funding from the

European Union's Horizon 2020 programme under the Marie Skłodowska-Curie grant agreement Ref. 843830. N.C.H.H. acknowledges funding from the European Union's Horizon 2020 programme under the Marie Skłodowska-Curie grant agreement Ref. 665884. J.H.E. acknowledges support from the Office of Naval Research (award N00014-20-1-2474) for the hBN crystal growth. D.M-C. and V.P. acknowledges funding from Government of Spain's project TUNA-SURF (PID2019-106892RB-I00) and from Generalitat de Catalunya's project AGAUR 2017 SGR 1634. D.M-C also wishes to acknowledge support from Ayuda (PRE2017-082781) and from FSE project "El FSE invierte en tu futuro". M.J. and G.S. acknowledge the support by the Office of Naval Research (ONR) under Grant No. N00014-21-1-2056, and by the National Science Foundation (NSF) under Grants No. DMR-1741788 and DMR-1719875. M.J. was also supported in part by the Kwanjeong Fellowship from the Kwanjeong Educational Foundation.

Data availability:

Relevant data supporting the key findings of this study are available within the article and the Supplementary Information file. All raw data generated during the current study are available from the corresponding author upon request.

Contributions:

H.H.S., D.M.C., L.O., M.C., and D.B.R., worked on sample fabrication. Isotopic hBN crystals were grown by E.J. and J.E. Measurements were performed by H.H.S. with help from L.O. and M. C. Numerical modelling was performed by M.J., A.M. and G.S. Experiments were designed by H.H.S., L.O., and F.H.L.K., with help from I.T., M.C. and G.S. All authors contributed to manuscript writing. J.E., V.P., G.S. and F.H.L.K. supervised the work.

# Supplementary material for "Transverse hypercrystals formed by periodically modulated phonon-polaritons"

**Authors** Hanan Herzig Sheinfux[1], Minwoo Jung[2], Lorenzo Orsini[1], Matteo Ceccanti[1], Aditya Mahalanabish[2], Daniel Martinez-Cercós [1], Iacopo Torre[1], David Barcons Ruiz[1], Eli Janzen[3], James H. Edgar[3], Valerio Pruneri[1], Gennady Shvets[2], Frank H.L. Koppens[1,4]

[1]ICFO-Institut de Ciencies Fotoniques, 08860 Castelldefels (Barcelona), Spain

[2] School of Applied and Engineering Physics, Cornell University, Ithaca, New York 14853, USA

[3]Tim Taylor Department of Chemical Engineering, Kansas State University, Durland Hall, Manhattan, KS 66506-5102, USA

[4]ICREA-Institució Catalana de Recerca i Estudis Avançats, 08010 Barcelona, Spain

# Table of Contents

## S1. Hyperbolic phonon polaritons

In this section we briefly consider the PhP modes in hBN, which play an instrumental role in our work. We will follow the notation and derivation in [1] and [2] in order to obtain the dispersion of these modes and motivate the complex bandstructure shown in the main text.

We consider a hyperbolic dispersion medium slab, infinitely extended in the $x - y$ direction, defined by the permittivity:

$$\epsilon_{xx}(\omega; z) = \begin{cases} 1 & z > t \\ \epsilon_x & 0 < z < t \\ \epsilon_s & z < 0 \end{cases}$$
$$\epsilon_{zz}(\omega; z) = \begin{cases} 1 & z > t \\ \epsilon_z & 0 < z < t \\ \epsilon_s & z < 0 \end{cases}. \qquad [1]$$

Where $\epsilon_s$ is the substrate permittivity and the superstrate is assumed to be air. In the upper Restrahlen band $\epsilon_x(\omega)$ is a mostly real, negative function of the frequency, whereas $\epsilon_x$ is a positive mostly real number, practically independent of the frequency. From Maxwell equations, we can obtain (e.g. [1,2]) the following mode dispersion

$$q_n(\omega; \epsilon_s) = \theta \frac{\pi n + \rho}{t}. \qquad [2]$$

Where $q_n$ is the mode momentum in the $x$ direction, $\theta = \sqrt{-\frac{\epsilon_z}{\epsilon_x}}$ is a constant (note that the argument of the square root is positive, since $\epsilon_x, \epsilon_z$ have opposite signs), $t$ is the thickness of the hBN and $\rho = \log \frac{i\theta \epsilon_x - 1}{i\theta \epsilon_x + 1}$ is a complex number. Equivalently, we can find the PhP wavelength

$$\lambda_n(\omega; \epsilon_s) = \frac{\pi t}{\theta(\pi n + \rho)}. \qquad [3]$$

We can immediately notice that since $\rho \neq 0, \pi$ inside the Restrahlen band, $\frac{\lambda_n}{\lambda_{n+1}} \neq const$. That is, the polariton wavelength is not a harmonic of some basic length scale. Therefore, there is no native length scale to the problem, except for the periodicity, in accordance with the overloaded, almost uniform, bandstructure seen in the fully folded simulation (Fig. 2b of the main text).

In addition, we can also see that the momentum strongly depends on the identity of the substrate, anticipating strong reflection at the interface where the substrate changes from dielectric to metallic.

## S2. Finite element simulation
### S2.1 Self-oscillating Bloch state of BN Phonon Polariton

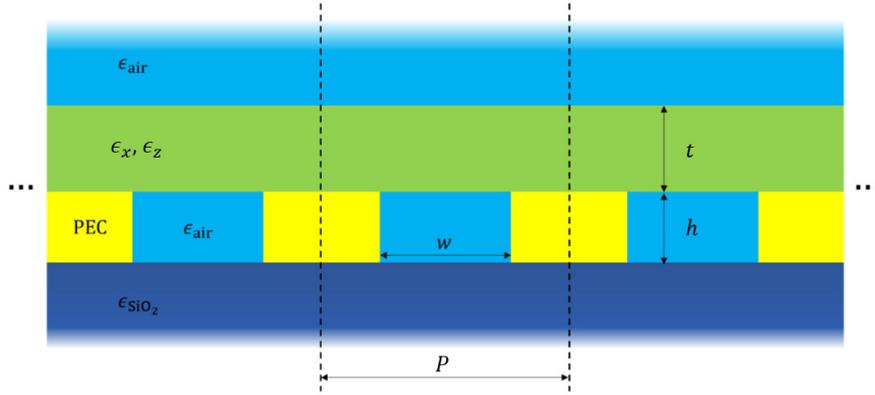

**Fig. S1** – Schematic of the simulated system, for a 1D periodic array of perfect electric conductor (PEC) strips, with the dimensions defined on the drawing.

Let $\mathbf{q}_n \equiv \mathbf{q} + \mathbf{G}_n$, where $n$ is an index counting the reciprocal unit vectors. For example, in 1D systems $\mathbf{G}_n = \hat{\mathbf{x}}\frac{2\pi}{L}n$. In 2D systems, $n$ refers to a number pair $n = (n_1, n_2)$ and $\mathbf{G}_n = \hat{\mathbf{x}}\frac{2\pi}{L_x}n_1 + \hat{\mathbf{y}}\frac{2\pi}{L_y}n_2$ (if the unit cell is a rectangle).

If the air regions (etched region in gold layer) are isolated from each other, we can write our ansatz as

$$\phi_{\mathbf{q}}(\mathbf{r},z) = \begin{cases} \sum_n e^{i\mathbf{q}_n \cdot \mathbf{r}} c_n^{\text{air}} e^{-|q|(z-h_1)}, & (z > h_1) \\ \sum_n e^{i\mathbf{q}_n \cdot \mathbf{r}} \left( c_n^{\text{BN}} \cos(\theta^{-1}|\mathbf{q}_n|z) + s_n^{\text{BN}} \sin(\theta^{-1}|\mathbf{q}_n|z) \right), & (0 < z < h_1) \\ \sum_\mu \sum_m e^{i\mathbf{q}\cdot\mathbf{R}_m} f_\mu^{\mathbf{q}}(\mathbf{r}-\mathbf{R}_m)\left( c_\mu^{\text{hole}} \cosh(\lambda_\mu^{\mathbf{q}} z) + s_\mu^{\text{hole}} \sinh(\lambda_\mu^{\mathbf{q}} z) \right), & (-h_2 < z < 0) \\ \sum_n e^{i\mathbf{q}_n \cdot \mathbf{r}} c_n^{\text{SiO}_2} e^{|\mathbf{q}_n|(z+h_2)}, & (z < -h_2) \end{cases}$$

[4]

Here, $m$ is an index counting the real space unit cells. For example, in 1D systems (see Fig. S1 above), $\mathbf{R}_m = mL\hat{\mathbf{x}}$; in 2D systems, $m$ refers to a number pair $m = (m_1, m_2)$ and $\mathbf{R}_m = m_1 L_x \hat{\mathbf{x}} + m_2 L_y \hat{\mathbf{y}}$ (if the unit cell is a rectangle). $c_n^{\text{part}}, s_n^{\text{part}}$, for part = air, BN, hole, SiO2, are coefficients. Then, $f_\mu^{\mathbf{q}}$ and $\lambda_\mu^{\mathbf{q}}$ satisfy the Laplacian eigenequation $\nabla_\mathbf{r}^2 f_\mu^{\mathbf{q}} + \lambda_\mu^{\mathbf{q}2} f_\mu^{\mathbf{q}} = 0$ with the Dirichlet boundary condition of $\lambda_\mu^{\mathbf{q}} f_\mu^{\mathbf{q}} = 0$ at the hole boundary ($\because E_z = -\partial_z \phi = 0$ at the lateral PEC surfaces). In the case of 1D periodic grating, they are given as

$$f_\mu^{\mathbf{q}}(\mathbf{r}) = \begin{cases} \sqrt{2}\sin\left(\frac{\mu\pi}{w}\left(x + \frac{w}{2}\right)\right) \text{ for } \mu \geq 1, \ 1 \text{ for } \mu = 0 & \left(-\frac{w}{2} < x < \frac{w}{2}\right) \\ 0, & (\text{otherwise}) \end{cases}, \quad \lambda_\mu^{\mathbf{q}} = \frac{\mu\pi}{w} \ (\mu \geq 0). \quad [5]$$

$f_\mu^{\mathbf{q}}$ is normalized so that $\int_{-\infty}^{\infty} f_\mu^{\mathbf{q}*} f_\nu^{\mathbf{q}} dx = w\delta_{\mu\nu}$. Similarly, for 2D systems, we can solve the Laplacian eigenequation for any arbitrary shapes of such isolated air holes to define $f_\mu^{\mathbf{q}}$ that vanishes at the

PEC boundary, and $f_\mu^q$ is normalized so that $\int_{-\infty}^{\infty} f_\mu^{q*} f_\nu^q d^2\mathbf{r} = S\delta_{\mu\nu}$ ($S$ is the area of the air hole in a unit cell).

On the other hand, if the air region (etched region gold layer) is a single connected hole (thus, having isolated islands of PEC) in 2D periodic systems, the ansatz for the PEC layer changes as

$$\phi_\mathbf{q}(\mathbf{r},z) = \sum_\mu f_\mu^\mathbf{q}(\mathbf{r})\left(c_\mu^{\text{hole}} \cosh(\lambda_\mu^\mathbf{q} z) + s_\mu^{\text{hole}} \sinh(\lambda_\mu^\mathbf{q} z)\right), \qquad [6]$$

where $f_\mu^q$ and $\lambda_\mu^q$ satisfy the Laplacian eigenequation $\nabla_\mathbf{r}^2 f_\mu^q + \lambda_\mu^{q^2} f_\mu^q = 0$ with the Bloch boundary condition $f_\mu^\mathbf{q}(\mathbf{r}+\mathbf{R}_m) = e^{i\mathbf{q}\cdot\mathbf{R}_m} f_\mu^\mathbf{q}(\mathbf{r})$ and the Dirichlet boundary condition of $\lambda_\mu^q f_\mu = 0$ at the hole boundary. Generally, this is not analytically solvable, so we obtained $f_\mu^q$ and $\lambda_\mu^q$ using the COMSOL Multiphysics Differential Equation module. $f_\mu^q$ is normalized so that $\int_{-\infty}^{\infty} f_\mu^{q*} f_\nu^q d^2\mathbf{r} = S\delta_{\mu\nu}$ ($S$ is the area of the air region in a unit cell).

Now we require the continuity of $\phi_q$ and the continuity of $\epsilon_z \partial_z \phi_q$. For simplicity, we use a compact vector notation

$$\begin{aligned}|\text{air}\rangle &= \{c_n^{\text{air}}\},\ |\text{cBN}\rangle = \{c_n^{\text{BN}}\},\ |\text{sBN}\rangle = \{s_n^{\text{BN}}\}, \\ |\text{ch}\rangle &= \{c_\mu^{\text{hole}}\},\ |\text{sh}\rangle = \{s_\mu^{\text{hole}}\},\ |\text{SiO}_2\rangle = \{c_n^{\text{SiO}_2}\}\end{aligned} \qquad [7]$$

and we introduce the matrices

$$[\hat{C}_1]_{n_1,n_2} = \cos(\psi|\mathbf{q}_{n_1}|h_1)\delta_{n_1,n_2}, \quad [\hat{S}_1]_{n_1,n_2} = \sin(\psi|\mathbf{q}_{n_1}|h_1)\delta_{n_1,n_2}, \quad [\hat{K}]_{n_1,n_2} = |\mathbf{q}_{n_1}|\delta_{n_1,n_2}$$

$$[\hat{C}_2]_{\mu_1,\mu_2} = \cosh(\lambda_{\mu_1}^q h_2)\delta_{\mu_1,\mu_2},\ [\hat{S}_2]_{\mu_1,\mu_2} = \sinh(\lambda_{\mu_1}^q h_2)\delta_{\mu_1,\mu_2},\ [\hat{\Lambda}]_{\mu_1,\mu_2} = \lambda_{\mu_1}^q \delta_{\mu_1,\mu_2}$$

$$\begin{aligned}[\hat{B}]_{\mu,n} &= \frac{1}{w}\int_{\text{in hole region}} f_\mu^{q*}(x) e^{i\mathbf{q}_n\cdot\mathbf{r}} dx \quad \text{(1D system)}, \\ &= \frac{1}{S}\int_{\text{in hole region}} f_\mu^{q*}(\mathbf{r}) e^{i\mathbf{q}_n\cdot\mathbf{r}} d^2\mathbf{r} \quad \text{(2D system)}\end{aligned}$$

$$\begin{aligned}[\hat{P}]_{n_1,n_2} &= \frac{1}{L}\int_{\text{in hole region}} e^{-i\mathbf{q}_{n_1}\cdot\mathbf{r}} e^{i\mathbf{q}_{n_2}\cdot\mathbf{r}} dx \quad \text{(1D system)}, \\ &= \frac{1}{L_x L_y}\int_{\text{in hole region}} e^{-i\mathbf{q}_{n_1}\cdot\mathbf{r}} e^{i\mathbf{q}_{n_2}\cdot\mathbf{r}} d^2\mathbf{r} \quad \text{(2D system)}\end{aligned} \qquad [8]$$

The matrix $\hat{B}$ is the transformation matrix from the plane waves to the Laplacian eigenfunctions inside the holes, and the matrix $\hat{P}$ is the projection matrix into the hole region.

There is a useful relation,

$$\hat{P} = \frac{w}{L}\hat{B}^\dagger \hat{B}, \qquad [9]$$

which can be derived using the completeness of the Laplacian eigenfunctions inside the holes: $\sum_\mu f_\mu^{q*}(\mathbf{r}) f_\mu^\mathbf{k}(\mathbf{r}') = \delta(\mathbf{r}-\mathbf{r}')$ (inside the hole).

Then, the continuity of $\phi_\mathbf{k}$ can be expressed as

$$|\text{air}\rangle = \hat{C}_1|\text{cBN}\rangle + \hat{S}_1|\text{sBN}\rangle$$

$$|\text{cBN}\rangle = \hat{P}|\text{cBN}\rangle,\ |\text{ch}\rangle = \hat{B}|\text{cBN}\rangle$$

$$\hat{C}_2|\text{ch}\rangle - \hat{S}_2|\text{sh}\rangle = \hat{B}|\text{SiO}_2\rangle,\ |\text{SiO}_2\rangle = \hat{P}|\text{SiO}_2\rangle \qquad [10]$$

and the continuity of $\epsilon_z \partial_z \phi_q$ can be expressed as:

$$-\epsilon_{\text{air}}|\text{air}\rangle = \epsilon_\perp \psi(\hat{C}_1|\text{sBN}\rangle - \hat{S}_1|\text{cBN}\rangle)$$

$$\epsilon_{\text{air}}\widehat{\Lambda}|\text{sh}\rangle = \hat{B}\epsilon_\perp \psi \widehat{K}|\text{sBN}\rangle$$

$$\epsilon_{\text{air}}\widehat{\Lambda}(\hat{C}_2|\text{sh}\rangle - \hat{S}_2|\text{ch}\rangle) = \hat{B}\epsilon_{\text{SiO}_2}\widehat{K}|\text{SiO}_2\rangle$$

$$\hat{C}_2|\text{ch}\rangle - \hat{S}_2|\text{sh}\rangle = \hat{B}|\text{SiO}_2\rangle, \quad |\text{SiO}_2\rangle = \hat{P}|\text{SiO}_2\rangle \qquad [11]$$

We can eliminate $|\text{air}\rangle$ and $|\text{sBN}\rangle$.

$$\hat{C}_1|\text{cBN}\rangle + \hat{S}_1|\text{sBN}\rangle = -\frac{\epsilon_\perp \psi}{\epsilon_{\text{air}}}(\hat{C}_1|\text{sBN}\rangle - \hat{S}_1|\text{cBN}\rangle), \qquad [12]$$

$$\Rightarrow \left(\frac{\epsilon_\perp \psi}{\epsilon_{\text{air}}}\hat{C}_1 + \hat{S}_1\right)|\text{sBN}\rangle = \left(\frac{\epsilon_\perp \psi}{\epsilon_{\text{air}}}\hat{S}_1 - \hat{C}_1\right)|\text{cBN}\rangle, \qquad [13]$$

Similarly, we eliminate $|\text{SiO}_2\rangle$ and $|\text{sh}\rangle$.

$$\epsilon_{\text{air}}\widehat{\Lambda}(\hat{C}_2|\text{sh}\rangle - \hat{S}_2|\text{ch}\rangle) = \hat{B}\epsilon_{\text{SiO}_2}\widehat{K}|\text{SiO}_2\rangle = \hat{B}\epsilon_{\text{SiO}_2}\widehat{K}\hat{P}|\text{SiO}_2\rangle = \hat{B}\epsilon_{\text{SiO}_2}\widehat{K}\frac{w}{L}\hat{B}^\dagger\hat{B}|\text{SiO}_2\rangle$$

$$= \hat{B}\epsilon_{\text{SiO}_2}\widehat{K}\frac{w}{L}\hat{B}^\dagger(\hat{C}_2|\text{ch}\rangle - \hat{S}_2|\text{sh}\rangle) \qquad [14]$$

$$\Rightarrow \widehat{\Lambda}(\hat{C}_2|\text{sh}\rangle - \hat{S}_2|\text{ch}\rangle) = \hat{R}(\hat{C}_2|\text{ch}\rangle - \hat{S}_2|\text{sh}\rangle), \qquad [15]$$

where $\hat{R} = \frac{\epsilon_{\text{SiO}_2}}{\epsilon_{\text{air}}}\frac{w}{L}\hat{B}\widehat{K}\hat{B}^\dagger$.

$$\Rightarrow (\hat{R}\hat{S}_2 + \widehat{\Lambda}\hat{C}_2)|\text{sh}\rangle = (\hat{R}\hat{C}_2 + \widehat{\Lambda}\hat{S}_2)|\text{ch}\rangle, \qquad [16]$$

Finally, we get,

$$|\text{cBN}\rangle = \hat{P}|\text{cBN}\rangle = \frac{w}{L}\hat{B}^\dagger\hat{B}|\text{cBN}\rangle = \frac{w}{L}\hat{B}^\dagger|\text{ch}\rangle = \frac{w}{L}\hat{B}^\dagger(\hat{R}\hat{C}_2 + \widehat{\Lambda}\hat{S}_2)^{-1}(\hat{R}\hat{S}_2 + \widehat{\Lambda}\hat{C}_2)|\text{sh}\rangle$$

$$= \frac{w}{L}\hat{B}^\dagger(\hat{R}\hat{C}_2 + \widehat{\Lambda}\hat{S}_2)^{-1}(\hat{R}\hat{S}_2 + \widehat{\Lambda}\hat{C}_2)\widehat{\Lambda}^{-1}\frac{1}{\epsilon_{\text{air}}}\hat{B}\epsilon_\perp \psi \widehat{K}|\text{sBN}\rangle = \frac{\epsilon_\perp \psi}{\epsilon_{\text{air}}}\hat{Q}\widehat{K}|\text{sBN}\rangle, \qquad [17]$$

where $\hat{Q} = \frac{w}{L}\hat{B}^\dagger(\hat{R}\hat{C}_2 + \widehat{\Lambda}\hat{S}_2)^{-1}(\hat{R}\hat{S}_2 + \widehat{\Lambda}\hat{C}_2)\widehat{\Lambda}^{-1}\hat{B}$.

To confirm that $\hat{Q}$ is Hermitian in the lossless limit, we note

$$\hat{Q} = \frac{w}{L}\hat{B}^\dagger(\hat{R}\hat{C}_2 + \widehat{\Lambda}\hat{S}_2)^{-1}(\hat{R}\hat{S}_2 + \widehat{\Lambda}\hat{C}_2)\widehat{\Lambda}^{-1}\hat{B} = \frac{w}{L}\hat{B}^\dagger(\hat{R}\hat{C}_2 + \widehat{\Lambda}\hat{S}_2)^{-1}(\hat{R}\hat{S}_2\hat{C}_2 + \widehat{\Lambda}\hat{C}_2^2)\hat{C}_2^{-1}\widehat{\Lambda}^{-1}\hat{B}$$

$$= \frac{w}{L}\hat{B}^\dagger(\hat{R}\hat{C}_2 + \widehat{\Lambda}\hat{S}_2)^{-1}\left((\hat{R}\hat{C}_2 + \widehat{\Lambda}\hat{S}_2)\hat{S}_2 + \widehat{\Lambda}\right)\hat{C}_2^{-1}\widehat{\Lambda}^{-1}\hat{B}$$

$$= \frac{w}{L}\hat{B}^\dagger\left(\hat{S}_2 + (\hat{R}\hat{C}_2 + \widehat{\Lambda}\hat{S}_2)^{-1}\widehat{\Lambda}\right)\hat{C}_2^{-1}\widehat{\Lambda}^{-1}\hat{B}$$

$$= \frac{w}{L}\hat{B}^\dagger\left(\hat{S}_2\hat{C}_2^{-1}\widehat{\Lambda}^{-1} + (\hat{C}_2\hat{R}\hat{C}_2 + \hat{C}_2\widehat{\Lambda}\hat{S}_2)^{-1}\right)\hat{B}. \qquad [18]$$

Therefore, since $\hat{C}_2, \widehat{\Lambda},$ and $\hat{S}_2$ are diagonal and $\hat{B}$ and $\hat{R}$ are Hermitian, $\hat{Q}$ is also Hermitian.

Finally, we get

$$\left[\left(\frac{\epsilon_\perp \psi}{\epsilon_{\text{air}}}\hat{C}_1 + \hat{S}_1\right) + \left(\hat{C}_1 - \frac{\epsilon_\perp \psi}{\epsilon_{\text{air}}}\hat{S}_1\right)\frac{\epsilon_\perp \psi}{\epsilon_{\text{air}}}\hat{Q}\widehat{K}\right]|\text{sBN}\rangle = 0, \qquad [19]$$

For this to be a non-trivial solution, the matrix should be not invertible.

## S2.2 Reflectivity Formulation

Above, we obtained a complicated nonlinear-type eigenvalue problem, so rather than solving for eigenvalues and vectors, it is better to formulate polaritonic resonance as diverging reflectivity.

Adding a source term from the top, we have

$$\phi_{\mathbf{q}}(\mathbf{r},z) = \begin{cases} \sum_n e^{i\mathbf{q}_n \cdot \mathbf{r}}\left(c_n^{\text{air}} e^{-|\mathbf{q}_n|(z-h_1)} - s_n^{\text{air}} e^{+|\mathbf{q}_n|(z-h_1)}\right), & (z > h_1) \\ \sum_n e^{i\mathbf{q}_n \cdot \mathbf{r}}\left(c_n^{\text{BN}} \cos(\psi|\mathbf{q}_n|z) + s_n^{\text{BN}} \sin(\psi|\mathbf{q}_n|z)\right), & (0 < z < h_1) \\ \sum_m e^{i\mathbf{q}\cdot\mathbf{R}_m} \sum_\mu f_\mu^{\mathbf{q}}(\mathbf{r}-\mathbf{R}_m)\left(c_\mu^{\text{hole}} \cosh(\lambda_\mu^{\mathbf{q}} z) + s_\mu^{\text{hole}} \sinh(\lambda_\mu^{\mathbf{q}} z)\right), & (-h_2 < z < 0) \\ \sum_n e^{i\mathbf{q}_n \cdot \mathbf{r}} c_n^{\text{SiO}_2} e^{|\mathbf{k}_n|(z+h_2)}, & (z < -h_2) \end{cases}$$

[20]

Notably, $e^{+|\mathbf{q}_n|(z-h_1)}$ is not a physically realistic wave, since we are in the quasi static limit and the excitation momentum is very large compared to $k_0$. Rather, we can think of it as an impinging evanescent wave. Following the similar procedure taken above, we get:

$$|\text{cBN}\rangle = \frac{\epsilon_\perp \psi}{\epsilon_{\text{air}}} \hat{Q}\hat{K}|\text{sBN}\rangle$$

$$\begin{cases} |\text{air}\rangle - |\text{source}\rangle = \hat{C}_1|\text{cBN}\rangle + \hat{S}_1|\text{sBN}\rangle \\ -\epsilon_{\text{air}}(|\text{air}\rangle + |\text{source}\rangle) = \epsilon_\perp \psi\left(\hat{C}_1|\text{sBN}\rangle - \hat{S}_1|\text{cBN}\rangle\right) \end{cases}$$

$$\Rightarrow \begin{cases} |\text{cBN}\rangle = \left(\hat{C}_1 + \frac{\epsilon_{\text{air}}}{\epsilon_\perp \psi}\hat{S}_1\right)|\text{air}\rangle - \left(\hat{C}_1 - \frac{\epsilon_{\text{air}}}{\epsilon_\perp \psi}\hat{S}_1\right)|\text{source}\rangle \\ |\text{sBN}\rangle = \left(\hat{S}_1 - \frac{\epsilon_{\text{air}}}{\epsilon_\perp \psi}\hat{C}_1\right)|\text{air}\rangle - \left(\hat{S}_1 + \frac{\epsilon_{\text{air}}}{\epsilon_\perp \psi}\hat{C}_1\right)|\text{source}\rangle \end{cases}$$

[21]

Eliminating $|\text{sBN}\rangle$,

$$\left(\hat{C}_1 + \frac{\epsilon_{\text{air}}}{\epsilon_\perp \psi}\hat{S}_1\right)|\text{air}\rangle - \left(\hat{C}_1 - \frac{\epsilon_{\text{air}}}{\epsilon_\perp \psi}\hat{S}_1\right)|\text{source}\rangle$$
$$= \frac{\epsilon_\perp \psi}{\epsilon_{\text{air}}}\hat{Q}\hat{K}\left(\left(\hat{S}_1 - \frac{\epsilon_{\text{air}}}{\epsilon_\perp \psi}\hat{C}_1\right)|\text{air}\rangle - \left(\hat{S}_1 + \frac{\epsilon_{\text{air}}}{\epsilon_\perp \psi}\hat{C}_1\right)|\text{source}\rangle\right). \qquad [22]$$

Finally, we get

$|\text{air}\rangle =$

$$\left(\hat{C}_1 + \frac{\epsilon_{\text{air}}}{\epsilon_\perp \psi}\hat{S}_1 - \frac{\epsilon_\perp \psi}{\epsilon_{\text{air}}}\hat{Q}\hat{K}\left(\hat{S}_1 - \frac{\epsilon_{\text{air}}}{\epsilon_\perp \psi}\hat{C}_1\right)\right)^{-1} \left(\hat{C}_1 - \frac{\epsilon_{\text{air}}}{\epsilon_\perp \psi}\hat{S}_1 - \frac{\epsilon_\perp \psi}{\epsilon_{\text{air}}}\hat{Q}\hat{K}\left(\hat{S}_1 + \frac{\epsilon_{\text{air}}}{\epsilon_\perp \psi}\hat{C}_1\right)\right) |\text{source}\rangle. \qquad [23]$$

We define the total reflectivity matrix as

$$\hat{r}_{\text{tot}} = \left(\hat{C}_1 + \frac{\epsilon_{\text{air}}}{\epsilon_\perp \psi}\hat{S}_1 - \frac{\epsilon_\perp \psi}{\epsilon_{\text{air}}}\hat{Q}\hat{K}\left(\hat{S}_1 - \frac{\epsilon_{\text{air}}}{\epsilon_\perp \psi}\hat{C}_1\right)\right)^{-1} \left(\hat{C}_1 - \frac{\epsilon_{\text{air}}}{\epsilon_\perp \psi}\hat{S}_1 - \frac{\epsilon_\perp \psi}{\epsilon_{\text{air}}}\hat{Q}\hat{K}\left(\hat{S}_1 + \frac{\epsilon_{\text{air}}}{\epsilon_\perp \psi}\hat{C}_1\right)\right). \qquad [24]$$

Then, we can plot the loss function $-\Im(tr(\hat{r}_{tot}))$ as a function of **q** and $\omega$, so that the phonon-polariton appears as darker peaks in the band structure.

### S2.3 Single-folded and unfolded band structure

In the main text, we introduced a single folding of the band structure both as a way to visualize the bandstructure, which in the usual "fully folded" presentation is full of partially overlapping bands to the point of being overloaded, and also to better match the experimental conditions. Experimentally, light is coupled to PhPs in the nearfield via the sSNOM tip. This coupling is generally expected to become very inefficient when the relevant PhP wavelength is much smaller than the tip diameter. Therefore, the tip efficiency in coupling to various $\mathbf{q}_n$ components decreases with $n$ and for typical conditions direct coupling to modes with $n > 2$ is relatively small. Furthermore, due to absorption in the lattice, we can intuitively expect a limitation on the number of interactions with the lattice an initial excitation will experience, restricting the conversion from the first Brillouin zone to a high-order Brillouin zone.

All this justifies doing partial band-folding instead of full bandfolding. Intuitively, this can be thought of as restricting the number of lattice interactions. Mathematically, we take a subset of diagonal elements instead of taking the entire trace of $\hat{r}_{tot}$. Specifically, for the single folded bandstructure, we take the subset of the diagonal elements that correspond to the primary Bloch vector excitation/response, $[\hat{r}_{tot}]_{(0,0),(0,0)}$ (the red region in Fig. S2), and the wavevectors of the adjacent Brillouin zones (BZ), $[\hat{r}_{tot}]_{(-1,0),(-1,0)}$ (the orange region in Fig. S2), $[\hat{r}_{tot}]_{(0,-1),(0,-1)}$ (the yellow region in Fig. S2), $[\hat{r}_{tot}]_{(-1,-1),(-1,-1)}$ (the green region in Fig. S2). Thus, the single-folded band structure visualization was done by calculating $-\Im([\hat{r}_{tot}]_{(0,0),(0,0)} + [\hat{r}_{tot}]_{(-1,0),(-1,0)} + [\hat{r}_{tot}]_{(0,-1),(0,-1)} + [\hat{r}_{tot}]_{(-1,-1),(-1,-1)})$ over the boundary of irreducible BZ (marked as the thick black right triangle in Fig. S2).

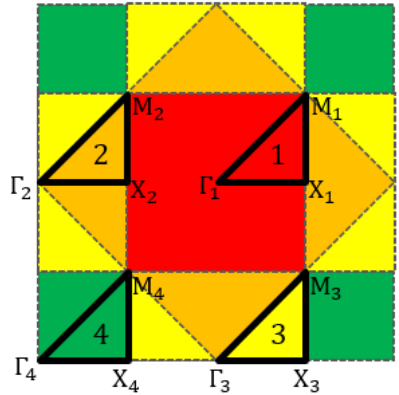

**Fig. S2** – The primary BZ (the red region marked as 1), secondary BZ (orange, 2), tertiary BZ (yellow, 3), and the quaternary BZ (green, 4).

Also, the radial average of the calculated band structure in Fig. S9, Fig. S14 was obtained by taking

$\int_{\mathbf{q}=q(\cos\theta,\sin\theta)} -\Im\left(tr([\hat{r}_{tot}(\mathbf{q},\omega)]_{(0,0),(0,0)})\right) d\theta.$

## S3. Sample fabrication

Key to our approach is the high resolution patterning of the metallic substrate, in contrast to patterning (e.g. etching, ion milling) the hBN directly. In doing so, we maintain high (pristine) hBN quality, whereas patterning the substrate with nanometrically precise FIB techniques, which are aggressive, would damage the hBN if used for direct patterning. At the same time, the contrast between the PhPs on the metallic substrate and on the dielectric substrate is very significant, generating strong reflection and intermodal coupling. The main ideas of our approach are explained in the main text and in the methods section of the main paper, but here we provide some additional technical detail.

In our experience, we find that the patterning of the gold structure can be performed with a variety of nanofabrication techniques, including electron beam lithography and $Ga^+$ FIB. However, $He^+$ and $Ne^+$ based fabrication seems to show superior results. Of these two techniques, He ion milling is more precise and more stable, but is also more prone to substrate-induced damage. Namely, ion backscattering which damages the milled pattern (as well as the substrate itself) and the deposition of He atoms inside the sample, which can induce potentially large topographical changes. These issues can be partially mitigated by using suspended ultrathin membranes, since the He ions pass through the membrane practically uninterrupted. The membranes are commercially available SiN membranes from Norcada with dimensions of $30\mu m$ or less and nominal thickness of 20nm. Gold is deposited on the membrane in the same process as for the chips.

While He FIB enjoys a superior resolution and current stability, we find Ne milling to be very well suited for structures with periods on the order of 25nm or more. The improved yield of Ne ions (compared to He) means that larger samples can be produced with less effect from mechanical drift. It also allows avoiding the use of suspended samples, which poses some problems for the measurement (see discussion in section 4). Regardless of the ion specie, patterning was designed using the NPVE patterning software.

For Ne milling, we used primarily lower currents ($0.5 - 2pA$), with a $10\mu m$ aperture and spot control of 4 or better. Ne pressure was set at $5\mu T$ and BIV voltage was increased above nominal value for Ne extraction to 88% of the He value or more. This stabilizes the current output, as well as improving source lifetime, at the expense of reducing the current magnitude. Extensive alignment to determine ideal beam conditions was performed in vicinity to the final patterned location. Demo structures were patterned and imaged in vicinity to the sample, but the final sample was not measured directly to avoid damage.

Following this, we transfered an isotopically pure hBN flake over the pattern using a variation of the PDMS based dry transfer technique. Exfoliation was done directly with low retention PDMS (X0, DGL type from Gelpak). The original crystal is exfoliated until the PDMS is visibly covered in hBN all around, at which point a fresh PDMS sheet was used to pick up a portion of the hBN on the original PDMS and further exfoliation can be performed. After 2-5 such rounds, a last round of exfoliation was done directly on the stamp. Flake thickness can be estimated by calibrating the optical contrast with AFM measurements of dropped flakes.

To drop the flakes, the chip was heated to $60°C$ and the PDMS is slowly brought into contact with the chip. After the hBN flake was fully brought into contact with the chip, we heated further to $85°C$. The PDMS stamp is then lifted extremely slowly and the hBN remains attached to the substrate by van der Waals forces. The flake was further cleaned from any transfer process residues by soaking in Acetone and isopropanol and, if needed, by contact mode AFM brooming.

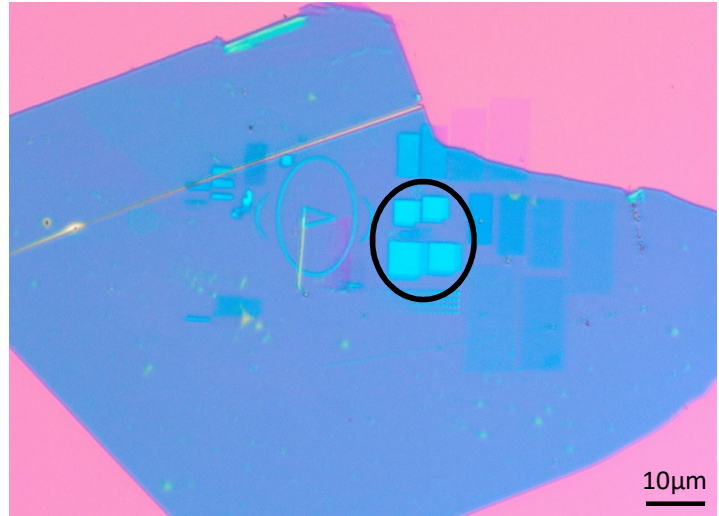

**Fig. S3** – The lattice device in an optical microscope, showing a set of hypercrystal devices. The blue shaded area is the isotopically pure hBN and the lighter shaded shapes are 4 lattices milled into the gold.

## S4. Nearfield measurement and analysis

The sSNOM is a widely used and commercially available technique that combines the accuracy of atomic force microscopy with optical excitation and measurement of deep subwavelength polaritons. All of our measurements have been performed using a SNOM (Neaspec) with standard atomic force microscope tips coated with a PtIr layer and with a nominal diameter of 40-50nm. All reported signals were extracted from $4^{th}$ harmonic measurements and the majority of the measurements have been performed in homodyne mode, which was very robust and less affected by laser frequency changes.

More specifically, frequency sweeps such as those shown in Fig. 4a are made by repeatedly scanning the same line at a sequence of laser frequencies. Experimentally, we find it preferable to normalizing relative to the $4^{th}$ harmonic signal outside of the lattice region (instead of to the laser power, for example). This type of measurement is naturally very sensitive to piezo drift, background effects and optical absorption, which depend in turn on factors such as lab humidity. For the measurements in Fig. 4a, the sSNOM was placed in a partial enclosure and flushed with nitrogen. Nevertheless, obtaining high quality frequency sweep measurements consistently proved difficult. For example, similar measurements of device HSQ1 showed more significant noise.

Regarding the extraction of the collective mode wavelength from single frequency pictures, which was performed in Fig. 3c, Fig. S9b and Fig S4. In many nearfield studies of polaritons, the polariton momentum can be extracted directly from the measurement by fitting to simple models and\or by applying functional transforms. In our case, we found such practice difficult and inaccurate for most frequencies of interest, mainly due to the complicated landscape. Collective modes coexist with features such as the spatial distribution of the Bloch wavefunction. Typically, for relatively long wavelengths (e.g. device HSQ1 regime I) the collective mode was very pronounced and long lived, hence the Bloch wavevector can be extracted by fitting. However, in almost all cases, the collective modes algorithmic fitting proved to be very challenging. Instead, we found it necessary to extract the collective mode wavelength manually, by evaluating the average fringe spacing from 2D scans and\or linetraces in a systematic fashion. This manner of evaluation surely introduces a degree of error, because the fringe spacing is also affected by geometrical spreading and absorption for example. Nevertheless, there is a notable degree of agreement between the extracted Bloch wavevectors and the theoretical calculations (e.g. Fig. 3c of the main text and Fig. S9b here), which indicates the manual extraction is reasonable accurate.

An important source of inaccuracy is the question of whether the mode is reflected from the defect\lattice edge or launched by it (and picked up by the sSNOM tip). If reflected, the distance between fringes should be $\lambda_{PhP}/2$, where launching means that a $\lambda_{PhP}$ distance is expected. In a few cases both launched and reflected components exist simultaneously and the resulting beating pattern can be identified. However, for the most part, we determine whether launching or reflection takes place by comparison to the calculated collective mode wavelength. As far as we can determine, in device NSQ1 and NSQ2, the fringes associated with the structural defect are always associated with polaritonic reflection, whereas the fringes associated with edge reflection change from mostly-launched to mostly-reflected depending on the frequency. See also Fig. S12 below and discussion surrounding that figure.

Regarding the interpretation of the signal, the nearfield interaction of the sSNOM tip with the sample can be modeled as an oscillating dipole source, oriented in the out of plane direction (e.g. reference[3]). In that approximation, the sSNOM signal is proportional to the zz component of the Green function of the electromagnetic field, the imaginary part of which is proportional to the z-projection of the local optical density of states (LDOS). In our homodyne measurement, the real and imaginary parts are mixed. Since the real and imaginary parts are tied via Kramers-Kronig relations, a sharp peak in the real part will be accompanied by a sharp response in the imaginary part and vice versa. With homodyne measurements, we cannot extract the exact frequency and bandwidth of an LDOS peak, but we can surmise that a sharp nearfield sSNOM response indicates a sharp LDOS peak occurs at a nearby frequency and with comparable width.

## S5. Additional measurements and calculations

To supplement the data in the main text, we present below additional measurements for the devices discussed in the text, in the order at which they appear. In addition, we also present data taken from additional devices. Namely, device NGR2, showing particularly narrow bandwidth density of states (DOS) peaks and device NSQ3, showing signal peaks that have no single mode equivalent, further cementing our claim on the multimodal nature of the bandstructure.

### S5.1 Device NSQ1

This device showed especially clear polaritonic fringes, for example, as seen in Fig. 1e of the main text and for a few additional frequencies, in Fig. S4 below. Curiously, the polaritonic response in this device was radially symmetric, in similarity to device NSQ2.

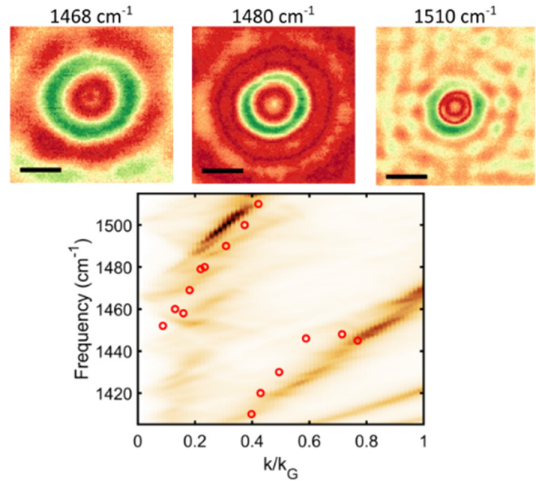

**Fig. S4 – Top**, 4th harmonic homodyne measurements of device NSQ1, taken near the structural defect at designated frequencies. Scalebar is 200nm. **Bottom**, collective mode momenta extracted from 2D scans on the background of the radial average of the calculated bandstructure.

The dispersion of collective modes momentum was extracted in a similar manner to Fig. 3c, except that the polaritons exhibited radial symmetry. The result, in Fig. S4, shows similar features to device NSQ2 (Fig. S9), except there is no psuedobandgap feature and the agreement with the theory is not as good, especially for the higher momentum branch. Device NSQ1 does, in addition, show a pronounced flat band where the collective mode momentum changes rapidly in a short frequency range, but unfortunately the device was accidently destroyed before further investigation of this feature could be confirmed.

### S5.2 Device NGR1

As explained in the main text, we also fabricated and measured an ultrafine lattice device, with a period of 25 nm. Both measurement and fabrication become very challenging on the 25nm scale. On the fabrication side, proximity effects and substrate damage restrict our ability to make disorder-free ultrafine lattices. On the measurement side, the lattice pitch is about 2.5 times smaller than the size of the sSNOM tip used in our measurement. This complicates measurements, reducing the strength of coupling by the tip to PhPs. Nevertheless, the lattice momentum in this device is about 280 times smaller than the mid-IR excitation wavelength, which is a record achievement for mid-IR sSNOM. Below, we show the 2D scans of device NGR1.

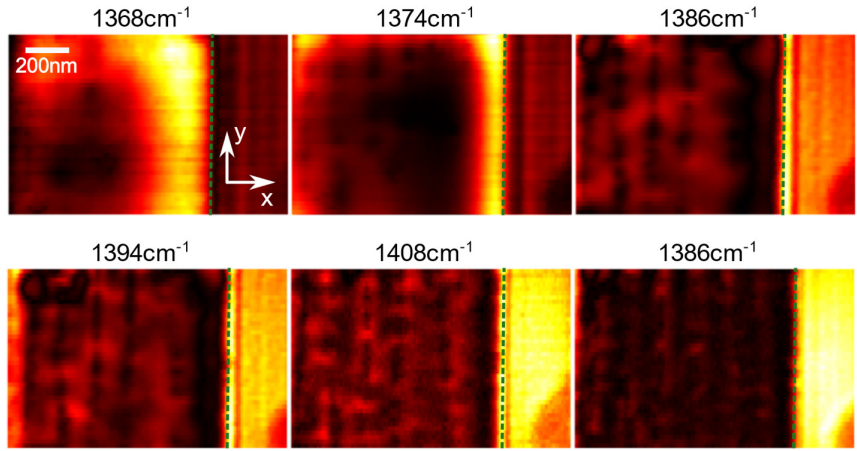

**Fig. S5** – sSNOM scans of device NGR1 at a range of frequencies. The lattice vector is oriented along the x-direction. The scalebar in the top left applies to all subfigures. The dashed green line indicates the position of the right lattice edge. As the frequency increases, we see the collective mode wavelength decreasing. Polaritonic fringes also appear outside of the lattice area, and are especially visible for lower frequencies and have a pronouncedly shorter wavelength than the collective mode. The colored patch on the bottom right of all subfigures is a bubble of polymeric residue.

As evident from these scans, this device suffers from some disorder and damage, damage which was not apparent in devices with larger periods. This damage is a consequence of proximity effects in the fabrication of these extremely dense lattices. To qualify the periodicity of the signal better, we perform spatial averaging of the signal. In the x-direction (parallel to the lattice vector) we show a range of frequencies below, with the averaging performed over almost the entire y-dimension. We also compare to the line trace in the y-direction (perpendicular to the lattice vector), in which case the average is taken over a 200nm wide strip since the quality of the edges parallel to the x direction is poorer and they are not uniform over a large area.

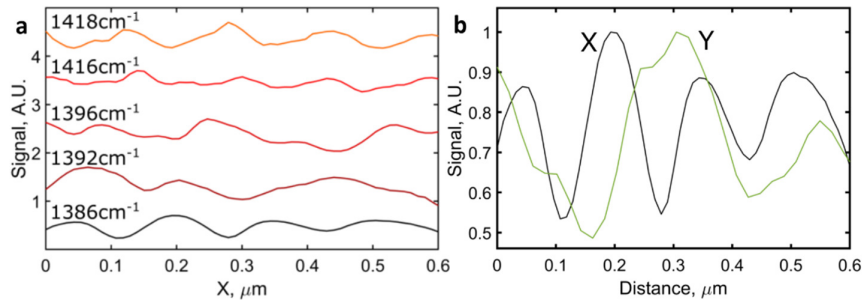

**Fig. S6** – **a.** Signal along a line extending from the right edge of the lattice, averaged in the y-dimension, at the frequencies indicated on the plot. **b.** The same, with the black line showing the signal at 1386 cm$^{-1}$ along the x-direction (along the lattice vector) and the green line showing the signal in the y-direction (perpendicular to the lattice vector). The ratio of the wavelengths in both directions is ~1.6.

### S5.3  Device HSQ1

The measurements of device HSQ1 shown in Fig. 2 of the main text were modified from the raw signal in two ways. First, a saturated colormap was used to make the measurement more visual, because the signal inside the lattice in this device was low relative to the exterior and. Second, for clarity, only a section of the lattice was shown in the main text Fig. 2. Below, we show the full lattice and using an unaltered colormap.

Notably, device HSQ1 exhibited a background signal, practically independent of the scan parameters (tapping amplitude, setpoint, etc.) and measurement technique (homodyne or pseudo-heterodyne). This signal is attributed to the mechanical oscillation of the ultrathin membrane induced by interaction with the tip. In our experience, relatively large membranes (50-100μm or more) show detrimental background artifacts, whereas smaller membranes (~20μm or smaller) do not. The particular membrane used for device HSQ1 was ~30μm large and shows pronounced artifacts, spots of higher\lower signal which vary on the μm-scale. This type of artifact was further complicated in pseudo heterodyne scans. The presence of the background signal also severely reduced the signal to noise ratio in frequency sweeps of this device. Nevertheless, the unsaturated plots (Fig. S7) shows clear periodic polaritons which are decreasing with frequency between panel a and panel b and between panel h and i (propoagating perpendicular to the diagonal). In panels d-f the fringe spacing is increasing with frequency (for example, compare bottom left of panel e and panel f). With a saturated colormap used in the main text this movement is easier to spot.

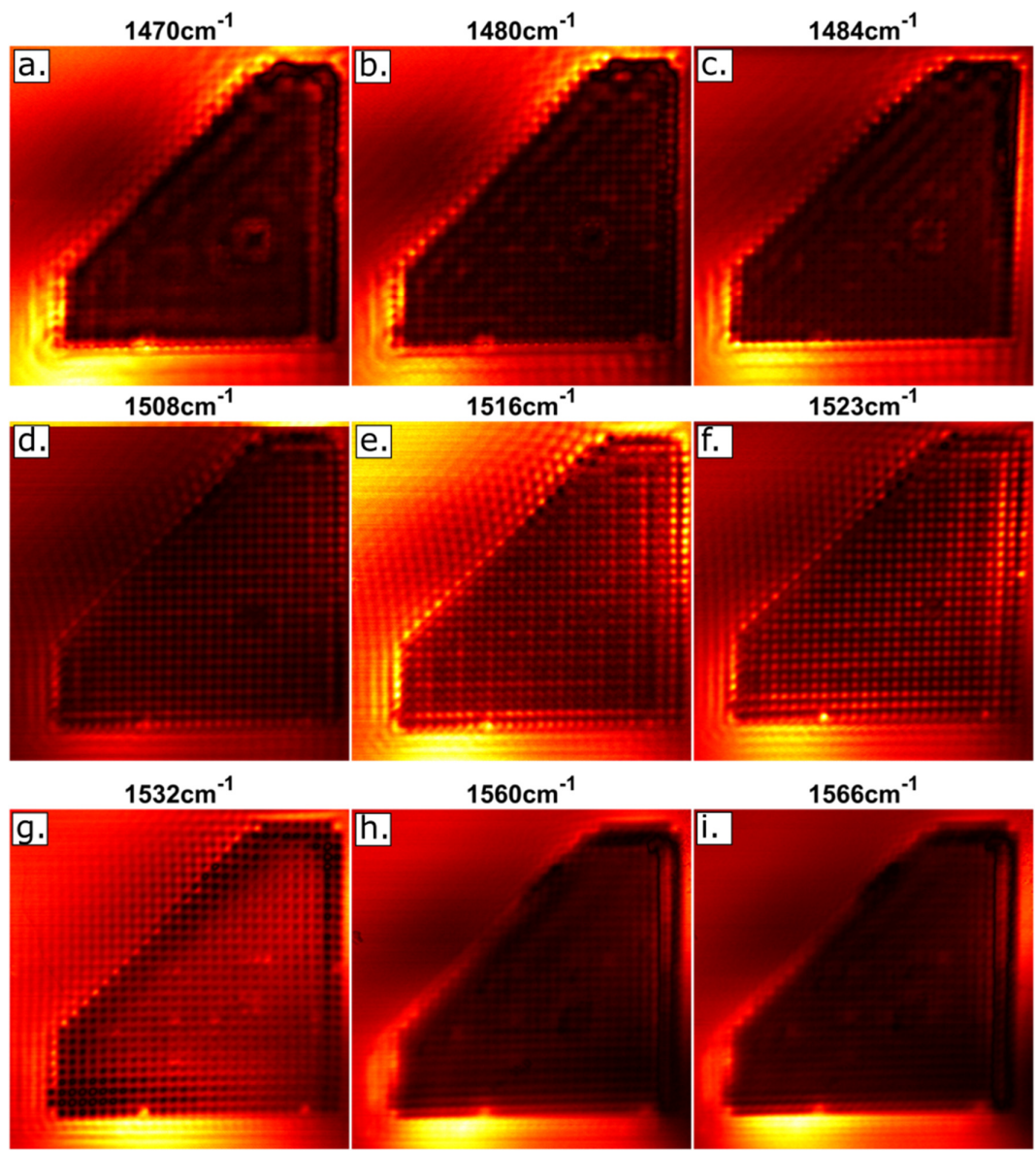

**Fig. S7** – Nearfield scans of device HSQ1, at the frequencies designated above each sub-panel. These figures are plotted with a linear unsaturated colormap (black to white is min to max). All scans show a 4.1µm area in width and length.

A notable feature of the above figure is that the decay length in panels a-c (regime II) is comparable inside and outside of the lattice. According to our best estimate, the decay length in this case is dominated by geometrical decay (the radial spread of PhPs away from the sSNOM tip) and not by absorption.

In addition to the above measurements, the bandstructure shown in the main text's Fig. 3a,c was also modified for visual clarity. This includes renormalization of the amplitude of the plot per frequency, the use of a nonlinear colormap and other cosmetic modifications. This is done to enhance the contrast and visibility of this energy loss plot, while keeping, to the best of our ability, the content of the figure unaltered. For completeness, we include below a version of the bandstructure with a linear colormap and without any modification. Notably, the unfolded bandstructure (Fig. 3b) used a modified colormap but was not renormalized.

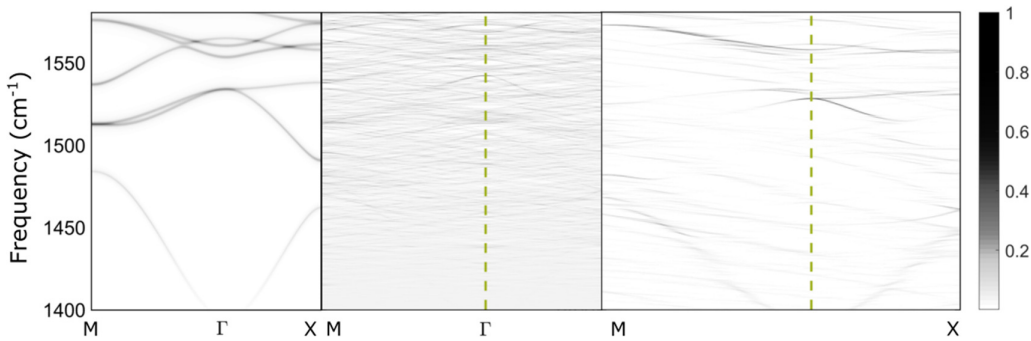

**Fig. S8** – Bandstructure of device HSQ1 in the single mode approximation (left), full calculation with full folding (middle) and single folding (right), without any cosmetic manipulation.

### S5.4 Device NSQ2

Similarly to what was done in Fig. 3 in the main text for device HSQ1, we produce below the bandstructure for device NSQ2. Specifically, we show the single mode bandstructure, the bandstructure without any folding and the radially averaged bandstructure, superimposed with the experimental results. The fully folded bandstructure for this device is not shown here, as it is similar to the fully folded bandstructure of other devices, densely populated with multiple modes. The single mode bandstructure is also not shown, as the zero folded bandstructure reproduces the experimental results more faithfully for this particular case.

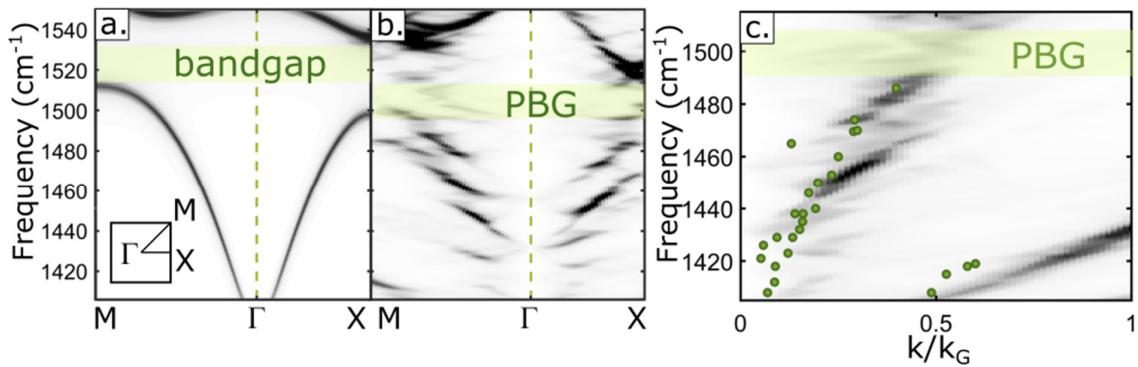

**Fig. S9** – **a.** Calculated bandstructure of device NSQ2, in the single mode approximation. **b.** Realistic bandstructure of device HSQ1, without any folding. **c.** Collective mode momenta extracted from 2D

scans above the radial average of the calculated bandstructure, as a function of $k/k_G$ with $k_G$ being the lattice wavevector. In all three subplots, a green shade highlights the location of the bandgap (single mode model) or pseudo-bandgap (multimodal case).

In contrast with device HSQ1, the Bloch modes in device NSQ2 are practically radially symmetric and are not oriented along the $\Gamma - X$ or $\Gamma - M$ axis. This symmetry is an unexpected finding, which warrants future investigation and is not clearly anticipated from theory. Similar radial symmetry was also seen in device NSQ1 and in other devices with relatively small metallic nanoisland size. Given this radial symmetry, we superimpose the experimentally extracted PhP momenta on top of radially averaged momentum plots.

As explained for device HSQ1, the bandstructure plots in Fig. S9 are modified for clarity, except for the radially averaged plot which did not require any such cosmetic modification. The unmodified originals are presented below.

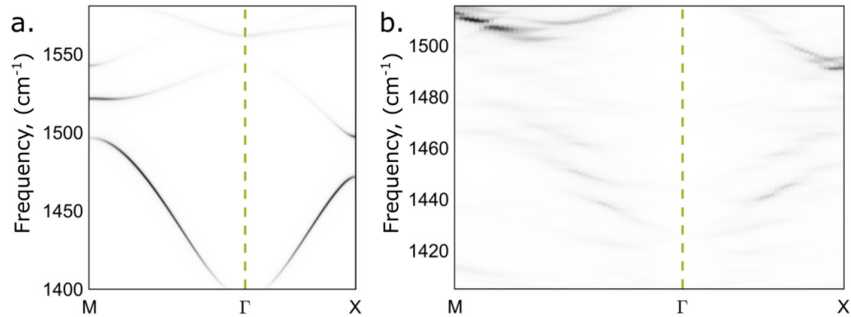

**Fig. S10** – Bandstructure of device NSQ2 in **a.** the single mode approximation and **b.** fully unfolded, without any cosmetic enhancement.

Below, we show a few additional single-frequency scans of device NSQ2 supplementing those in the main text.

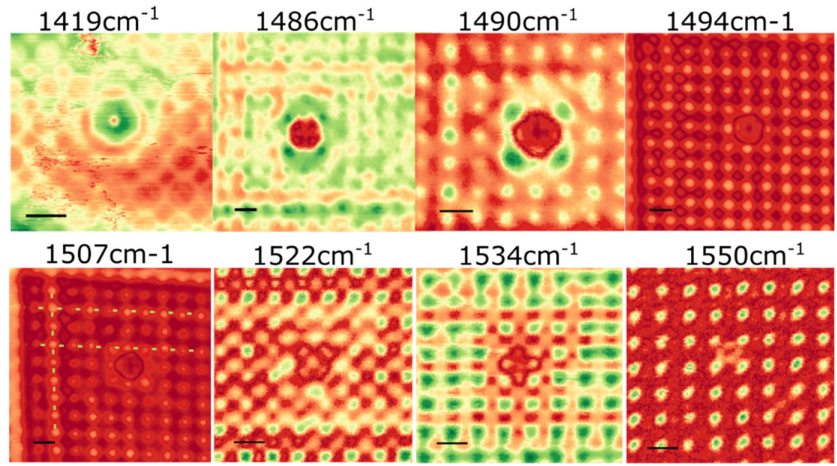

**Fig. S11** – Nearfield scans of device NSQ2, in the vicinity of a structural defect. Scans are at the frequencies indicated on the plot, focusing on frequencies around the PBG. Scale bars are 200nm in all plots. The green dashed line on the bottom left plot is a guide to the eye, showing the probable location of new polaritonic fringes above the PBG.

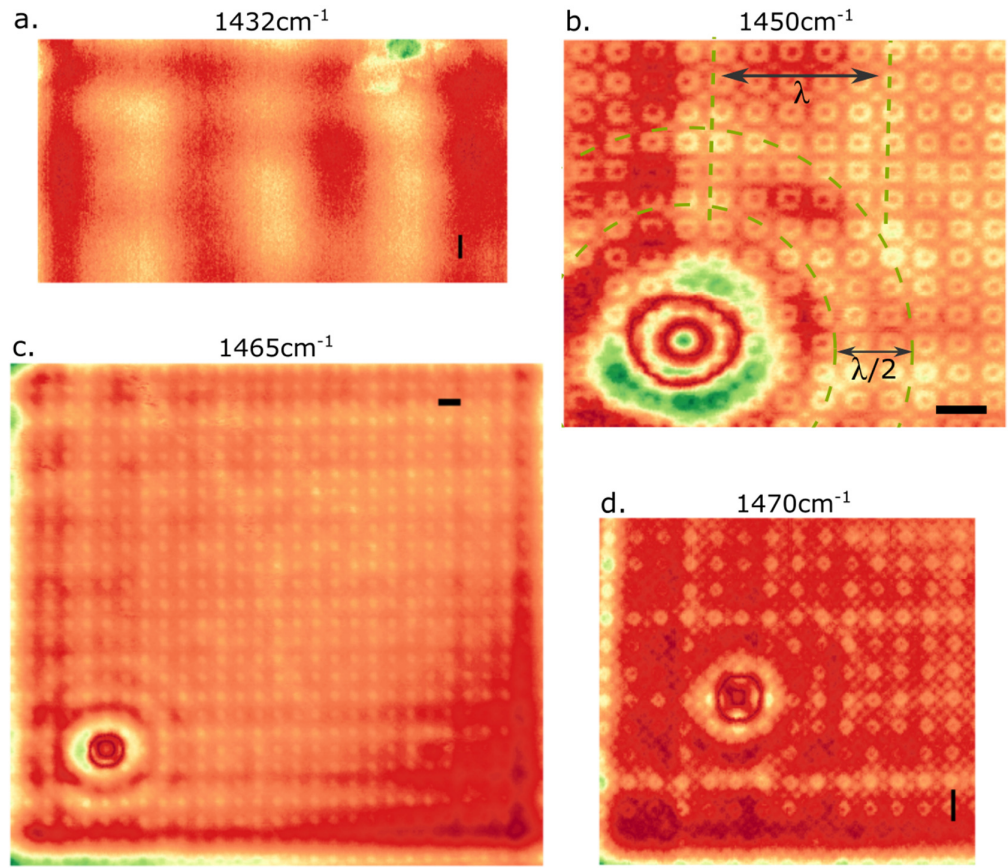

**Fig. S12** – Nearfield scans of device NSQ2, showing a larger area at the frequencies indicated on the plot. Scale bars are 200nm in all plots. **a.** Very long wavelength collective mode which forms a standing wave in the horizontal direction (green patch on top-right of the plot is due to dirt). **b.-d.** demonstrate the simultaneous appearance of fringes due to polaritons launched from the lattice edge and reflected polaritons (with half the period) emanating from the defect. Dashed green lines and circles are guides for the eye. The non-sine-like character of the fringes in **d** is similar to the fringes seen in regime III of device HSQ1 (see Fig. 2 of main text).

### S5.5  Device NGR2

This device is a 1D lattice (grating) with $p = 300$nm, $w = 200$nm, $t=15$nm, made with h$^{10}$BN. It is notable for showing extremely narrow signal peaks, on the order of 6 cm$^{-1}$, indicating that light is confined for a relatively long time on the nanoscale. We present below a frequency sweep similar to the one taken for device NSQ2 in Fig. 4a, as well as the averaged signal, similar to Fig. 4b.

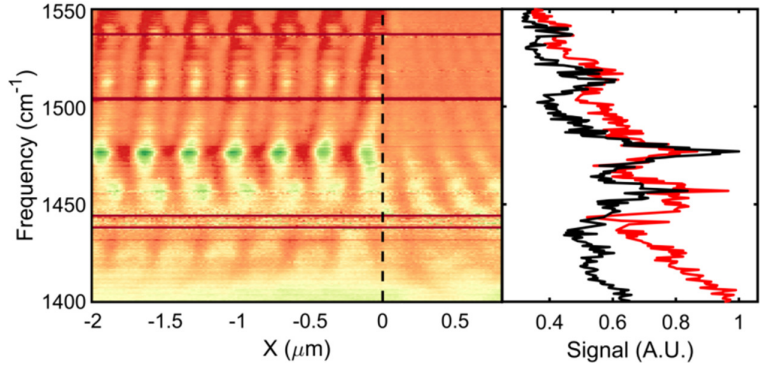

**Fig. S13** – Signal peaks in device NGR2 **a.** Frequency sweep, along the lattice vector direction. The dashed black line shows the edge of the lattice. **b.** The maximum signal (black line) and averaged signal (red line) measured in one unit cell, as a function of frequency.

The peaks in this device are exceptionally narrow. Namely, the pronounced peak near 1480 cm$^{-1}$ has a full width half maximum of about 6 cm$^{-1}$. To the best of our knowledge this is the narrowest such peak measured in any nanoscale device of comparable dimensions. Due to the complicated background of the signal and the homodyne measurement used to obtain it, this bandwidth cannot be translated accurately to a quality factor, but the quality factor associated with this peak is clearly appreciable. We note that agreement with theory for this particular device is poor, most likely due to a deviation between the designed and actual milled structure and our preference not to extensively to overfit by adjusting the simulated structure parameters to obtain agreement with experimental data.

S5.6 Device NSQ4

Device NSQ4 is a square lattice device with $p = 210$nm, $w = 170$nm, $t=25$nm made with h$^{11}$BN. Similarly to NSQ3, this device shows a complicated structure of spectral peaks, including some that have no possible single mode equivalent. Below, we show the calculated theoretical DOS, the single mode bandstructure and the experimental accumulated sSNOM signal. The calculated DOS agrees with the experimental signal, except for an 10cm$^{-1}$ offset. The origin of this offset could be related to compressive strain[4] or subtle structural differences between the simulated and actual device (e.g. changes in the hBN thickness, radius of the metallic corners, etc.). Importantly, the single mode bandstructure for the same device shows no features that could correlate to all of the experimental peaks. We therefore did not attempt to optimize the simulated structure to correct for this rather small shift, as the main purpose of this device is to show the deviation from the single mode model.

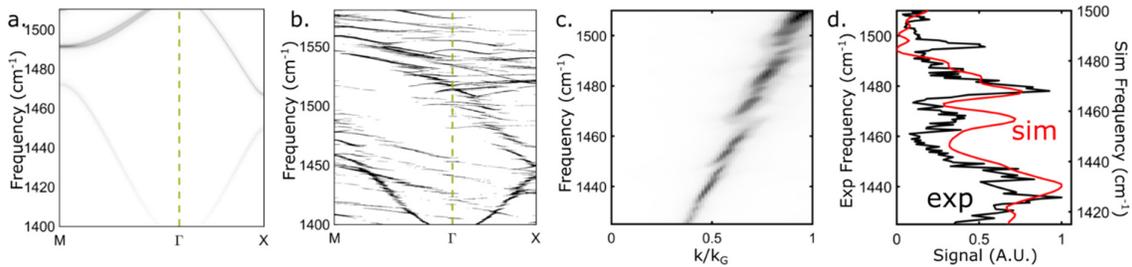

**Fig. S14** – **a.** A single mode model bandstructure for the same parameters as device NSQ4, showing no bandgaps open in the experimental range of frequencies. **b.** A single folded bandstructure plot is cosmetically enhanced for clarity. **c.** The radial average of the unfolded bandstructure **d.** The maximum experimental signal (black line) and theoretical DOS (red line), as a function of frequency. The experimental frequencies are shown on the left and are shifted by 10 cm$^{-1}$ relative to the frequency of the simulated signal. The experiment shows 4 strong peaks, of which the peaks at 1480 and 1500 cm$^{-1}$ correspond in spectral location to relatively flat bands in the bandstructure. The presence of these experimental peaks is in poor agreement with the single mode prediction.

## S6. Complexity in the fully folded bandstructure

One of the remarkable aspects of the theoretical calculation is the complexity of the bandstructure. The fully folded bandstructure (e.g. Fig. 3b of the main text) is practically overloaded with intersecting and interacting bands. As mentioned earlier, this complexity is even more notable because the geometrical simplicity of the lattice, where even a 1D lattice (a grating) can produce a complex array of DOS peaks.

While a through theoretical analysis of the emergent complexity is outside the scope of the current work, we do briefly return to the fully folded bandstructure in this section. To better appreciate the complexity of the bandstructure we plot the integrated density of states from the $\Gamma$ to the X of the fully folded bandstructrue of device HSQ2 (the same bandstructure shown in Fig. 3b of the main text). This partial DOS shows multiple narrow peaks and valleys, with typical widths on the sub $cm^{-1}$ scale. Notably, this calculation was done with a spectral resolution of 0.15 $cm^{-1}$, but the plot clearly is clearly not smooth and show many peaks and dips, some of which are a few $cm^{-1}$ and some appear to be too narrow to be fully resolved at this resolution. While many of these narrow DOS peaks are not expected to be experimentally accessible (due to limited tip coupling at high momenta), these plots are strikingly rich and quite clearly limited by the computational power used.

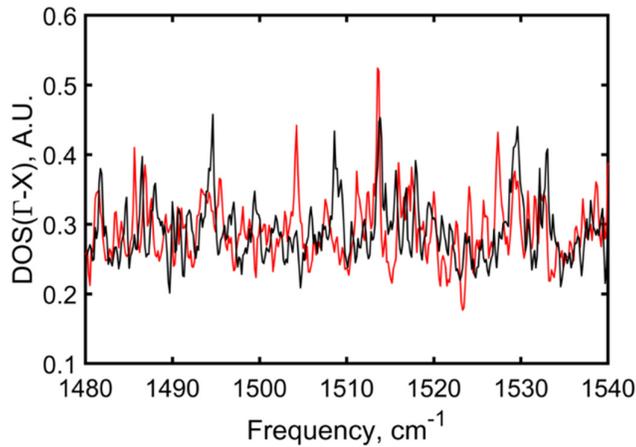

**Fig. S15** – The integrated signal of the fully folded bandstructure from the $\Gamma$ point to the X point. The red line shows the result for the parameters of device HSQ1. The black line shows a device with a slightly larger metallic island size (90 nm instead of 85 nm).

Also included in this plot is the calculated DOS in a device with a slightly different duty cycle (a 5nm larger metallic island size relative to a 120nm period). Notably, the slight shift in the duty cycle induces a significant variation in the calculated integrated density, demonstrating the important role of multimodal interaction. Consider the hypothetical case where the modes do not couple to each other. Reflection for higher order modes should be very small in this case at the metallic interface (because the impedance mismatch to the PhP modes on the metal decreases rapidly when the mode number increases) and the resulting bandstructure should be composed mostly of higher order modes which are artificially folded at the edges of the Brillouin zone. However, in that hypothetical scenario, changing the duty cycle would not change the bandstructure in any meaningful way (since most of the modes are barely reflected at the metal corners). The significant change in Fig S15 therefore indicates that multimodal reflection occurs even for very high mode numbers.